\pdfoutput=1

\documentclass[osajnl,twocolumn,showpacs,superscriptaddress,10pt]{revtex4-1} 
\usepackage{amsmath,amssymb,graphicx}
\usepackage{array}

\begin{document}

\title{Megawatt peak power from a Mamyshev oscillator}

\author{Zhanwei Liu}\email{Corresponding author: zl358@cornell.edu}
\author{Zachary M. Ziegler}
\author{Logan G. Wright}
\author{Frank. W. Wise}
\affiliation{School of Applied and Engineering Physics, Cornell University, Ithaca, New York 14853, USA}

\begin{abstract}
We demonstrate a fiber source with the best performance from an ultrafast fiber oscillator to date. The ring-cavity Mamyshev oscillator produces $\sim$ 50-nJ and $\sim$ 40-fs pulses. The peak power is an order of magnitude higher than that of previous lasers with similar fiber mode area. This performance is achieved by designing the oscillator to support parabolic pulse formation which enables the management of unprecedented nonlinear phase shifts. Experimental results are limited by available pump power. Numerical simulations reveal key aspects of the pulse evolution, and realistically suggest that (after external compression) peak powers that approach 10 MW are possible from ordinary single-mode fiber. The combination of practical features such as environmental stability, established previously, with the performance described here make the Mamyshev oscillator extremely attractive for applications.
\end{abstract}

\ocis{(060.2310) Fiber optics; (190.0190) Nonlinear optics; (060.4370)  Nonlinear optics, fibers.}

\maketitle 

\section{Introduction}

A consensus goal of research on ultrafast fiber lasers has been to develop an alternative to the solid-state mode-locked oscillator, with the purported benefits of the fiber platform: relatively low cost, simplicity, and robustness. Ultrafast lasers provide precision and high-intensity fields that have enabled many important advances in science.  Fiber ultrafast instruments could be transformative in enabling both widespread scientific and industrial applications of ultrafast pulses. However, for this they need to simultaneously reach sufficient performance and be amenable both to cost-effective manufacturing and use by non-experts. For a long time, the primary challenge to this end was the management of nonlinearity in the waveguide medium. In the past decade, this challenge has been met with several developments. New pulse evolutions based in normal dispersion fiber now provide a means of tolerating high nonlinearity \cite{Ilday04, Chong06,Chong20nj07, Nie11, Chong2015}. In laboratory prototypes that utilize nonlinear polarization evolution (NPE) as an effective saturable absorber, these sources rival solid-state oscillators. Their typical performance of $\sim$ 20-nJ and sub-100 fs pulses from standard single-mode fiber (SMF) represent order-of-magnitude higher peak power than early soliton \cite{Moll1984} and stretched-pulse fiber oscillators \cite{Tamura1993}. However, for widespread use and commercialization, NPE is unsuitable because it is highly sensitive to the random birefringence of the fiber, and consequently mode-locking is easily disrupted by environmental perturbations. This has become the impediment to the proliferation of fiber lasers in applications that employ femtosecond oscillators. It has stymied commercial instruments that reach beyond the scientific market.

Substantial effort has been devoted to solving this problem. Fiber lasers constructed with all polarization-maintaining (PM) fiber are robust against such environmental perturbations. To date, no work has been able to combine the high performance of NPE in standard SMFs with an all-PM design. Semiconductor saturable absorber mirrors (SESAMs) \cite{Chong08, Keller96} and nonlinear loop mirrors using PM fibers \cite{Aguergaray12, Aguergaray13, Szczepanek15} have been employed as alternative saturable absorbers. Material-based saturable absorbers, however, suffer from long-term reliability and poor power-handling capabilities. Nonlinear loop mirror (NOLM) and nonlinear amplifying loop mirror (NALM) based designs require precise control of the splitting ratio between loop directions, and their transmission cannot be easily and continuously tuned. These constraints make mode-locking in these lasers difficult to start. Careful and gradual adjustments are needed to reach the best performance \cite{Aguergaray13}, and the lasers exhibit a high sensitivity to pump fluctuations. Furthermore, although significant steps have been made, lasers based on SESAMs, NOLMs or NALMs still have not generated more than 5-nJ and sub-100 fs pulses. 

Devices based on reamplification and reshaping have been considered as an alternative for the generation of short pulses \cite{Pitois08, Sun09, North14}. This approach relies on self-phase-modulation (SPM) induced spectral broadening and offset spectral filtering, as proposed originally by Mamyshev for signal regeneration \cite{Mamyshev98}. These studies focus mainly on telecommunication applications, with the pulse energies and durations (usually picojoules and picoseconds) limited by nonlinear phase accumulation in long fibers and narrow filter separation \cite{Provost07}. Regelskis \textit{et al.} first demonstrated a Mamyshev oscillator aimed at high-energy femtosecond pulses \cite{Regelskis15}. This environmentally-stable oscillator produced modest-energy (\textless3 nJ) pulses about 2 ps in duration. The measured spectral bandwidth could support $\sim$150-fs transform-limited pulses, but the compressed pulse duration was not measured. These researchers also reported that the oscillator could be started by reflecting light rejected by the filter back into the cavity \cite{RegelskisPatent}. Very recently, starting of a Mamyshev oscillator by modulation of the pump laser at 20 kHz to induce Q-switching was reported \cite{Samartsev17}. This oscillator featured an all-fiber construction and generated impressive 15-nJ pulses, which were dechirped to 150-fs duration.

The recent works by Regelskis \textit{et al.} and Samartsev \textit{et al.} nicely illustrate the potential practical advantages of Mamyshev oscillators over conventional mode-locked lasers. On the other hand, they do not address the nature of the intracavity pulse propagation beyond the self-amplitude modulation that arises from the offset filtering. As a result, important questions remain regarding fundamental aspects of their operation. Additionally, the peak power obtained with such devices still lags behind that of mode-locked lasers that employ NPE. If Mamyshev oscillators cannot reach high performance levels, the benefits they provide over alternative environmentally-stable designs will be limited. On the other hand, if the performance limits of the Mamyshev oscillator meet or even exceed those of previous designs, the combination of performance and practical advantages may allow it to find widespread applications. Clearly, there is ample motivation to understand pulse propagation and its performance limits. Finally, the mechanisms of self-starting are still unclear.

Here, we report results of a theoretical and experimental study of pulse propagation in a Mamyshev oscillator. Insight gained about the pulse propagation allows us to achieve record performance for a femtosecond fiber oscillator, and forecasts ultimate performance limits yielding nearly an order-of-magnitude higher peak power. Numerical simulations show that an oscillator comprised of ordinary SMF, when seeded by low-energy short pulses (\textless10 ps), can support mode-locked pulses with 190-nJ energy and \textless20-fs dechirped duration (Fourier transform limited). These parameters correspond to $\sim$ 8 MW peak power. The simulations reveal that this performance follows from the oscillatoradds remarkable capacity to manage nonlinear phase. In a suitably-designed cavity, the pulse evolves to a parabolic shape before it enters the gain fiber, which enables control of the nonlinear phase accumulated in the gain segment. Theoretically, the pulse can accumulate a round-trip nonlinear phase shift of 140$\pi$ and still be stable. In experiments, $\sim$ 50-nJ and $\sim$ 40-fs pulses (after compression) are generated, with the energy limited by the available pump power and/or damage to the PM fiber. Even so, the $\sim$ 1 MW peak power is about 10 times higher than that has been produced by a mode-locked fiber laser constructed with ordinary SMF \cite{Chong20nj07}. The accumulated nonlinear phase shift is $\sim$ 60$\pi$, which is $\sim$ 5 times larger than the largest reported for stable pulses with well-controlled phase from a mode-locked laser. We attribute this to the parabolic pulse propagation along with the ``perfect'' saturable-absorber behavior of the Mamyshev process, which will be elaborated on below. These results represent a significant step toward a high-energy, short-pulse fiber source that can be environmentally-stable. Finally, we discuss fundamental aspects of the Mamyshev oscillator in the context of other driven nonlinear systems.

\begin{figure}[htbp]
\centerline{\includegraphics[width=1\columnwidth]{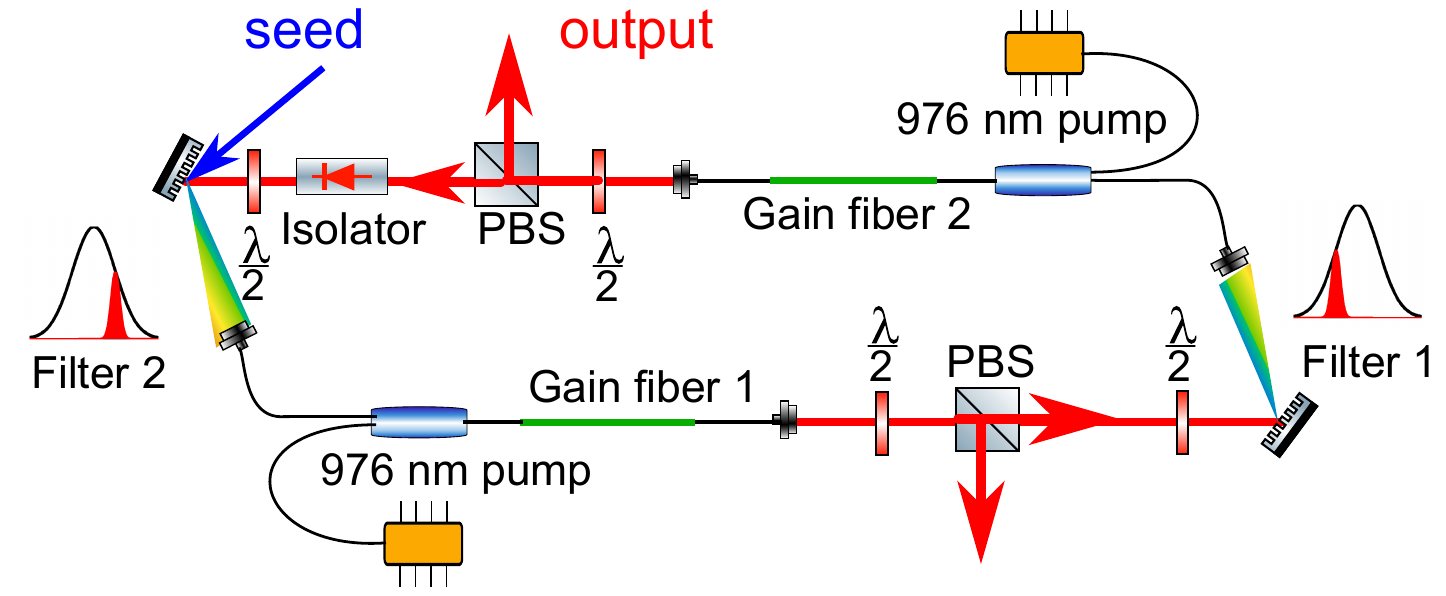}}
\caption{Schematic of the ring Mamyshev oscillator. Filter: The black curve shows the gain spectrum and the red curve indicates the passband of the filter; PBS: polarizing beam splitter.}
\label{labsetup}
\end{figure}

\section{Numerical and experimental results} 
The experimental configuration of the oscillator is schematically illustrated in Fig. \ref{labsetup}. The laser operates in the all-normal-dispersion regime (as do all prior Mamyshev oscillators). A ring oscillator allows more design freedom to control the propagation compared to a linear oscillator. Ytterbium-doped fiber provides the gain, and all fibers are PM. The use of Gaussian spectral filters is important for maximizing the pulse quality and peak performance (see Supplement 1, section 2). To accomplish this, we use the overlap of the beam diffracted from a grating with the spatial mode of the fiber \cite{Renninger2010}. These filters inherit their Gaussian shape from the fundamental mode shape of the SMF, and were tuned to longer ($\sim$ 1040 nm) and shorter ($\sim$ 1030 nm) wavelengths than the $\sim$ 1035 nm peak of the gain spectrum. An isolator ensures unidirectional operation. Polarizing beam splitters are used as the output couplers. The steady-state operation cycle consists of amplification (gain fiber 1), spectral broadening (gain fiber 1 and the following passive SMF), pulse energy adjustment (PBS), filtering (filter 1), amplification (gain fiber 2), spectral broadening (gain fiber 2 and the following passive SMF), output and spectral filtering (filter 2). The half-wave plates are used to adjust the polarization state of light going into the PM fiber and to optimize the output coupling ratio. To ignite pulsation in the cavity, a seed pulse is directed into the fiber via a grating. 

\begin{figure}[h!!]
\centerline{\includegraphics[width=1\columnwidth]{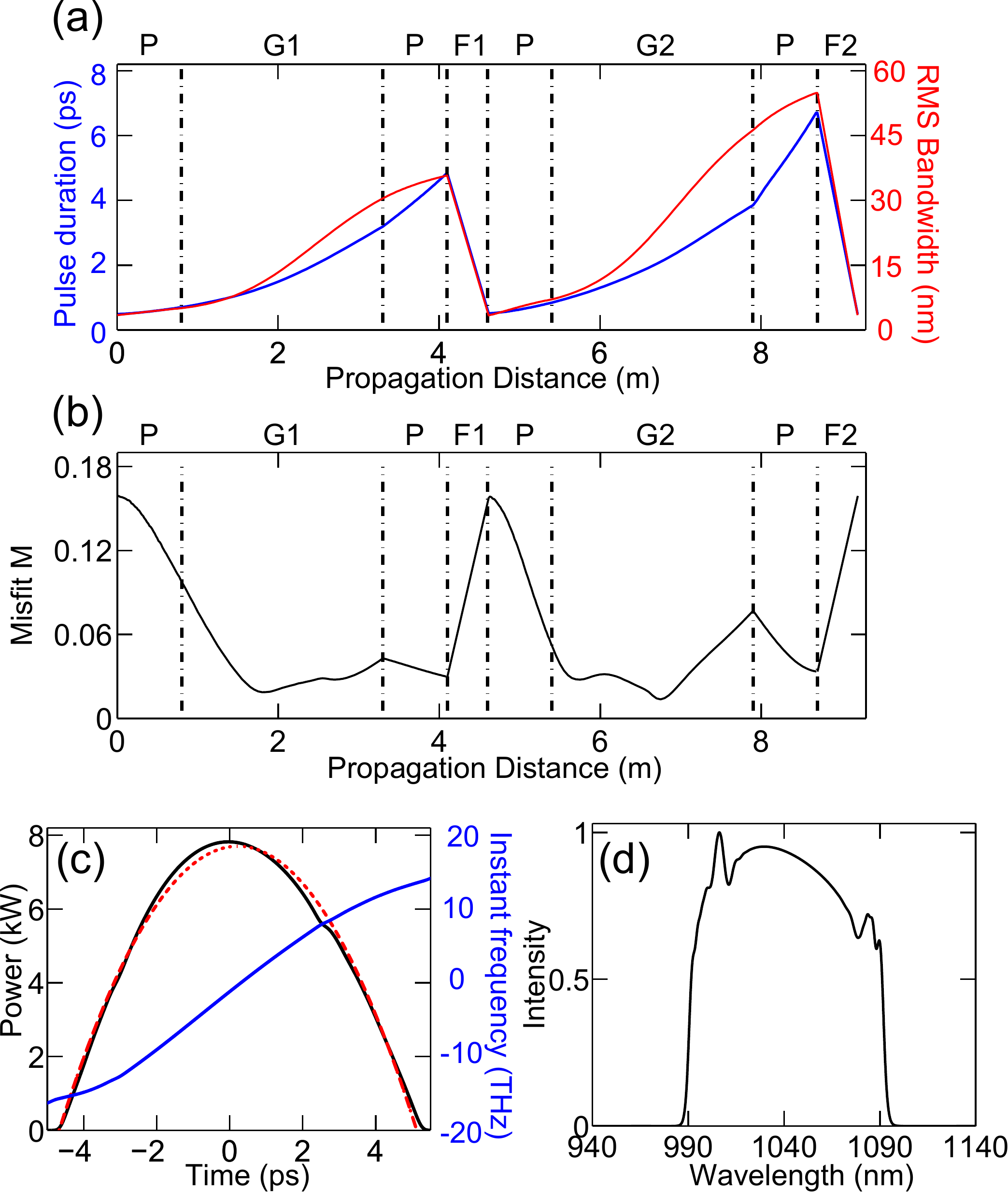}}
\caption{Numerical simulation results for $\sim$ 50 nJ output pulses. (a) Evolution of pulse duration (blue) and RMS bandwidth (red), P: passive fiber; G: gain fiber; F: filter; (b) evolution of misfit parameter M defined by $M^2 = \int (I-I_{fit})^2 dt / \int I^2 dt$, which indicates the difference between the pulse shape ($I$) and the best-fit parabolic profile $I_{fit}$; (c) temporal profile (black) with fitted parabolic curve (red) and instantaneous frequency across the chirped pulse (blue) and (d) simulated spectrum.}
\label{simulation}
\end{figure}

We performed numerical simulations of the oscillator shown in Fig. \ref{labsetup} using the standard split-step method with accurate fiber parameters. The simulation includes the Kerr nonlinearity, stimulated Raman scattering, and second and third order dispersion. The oscillator is seeded with different initial pulses (picosecond or femtosecond duration), but for given cavity parameters the simulations always converge to the same solution. The simplified gain dynamics and small temporal window of our simulations prevent our current numerical model from providing realistic insight into starting of pulse-formation from noise. Numerical simulations predict the cavity may generate up to 190-nJ pulses, which can be dechirped to below 20 fs. The pulse energy is limited by deviations of the pulse from a parabolic shape, which causes wave breaking, and by stimulated Raman scattering (see Supplement 1, Fig. 1). Another example, for which 50-nJ pulses are produced, is shown in Fig. \ref{simulation} for comparison to experimental results below. The pulse duration and bandwidth grow monotonically in the passive (80 cm), gain (2.5 m) and second passive fiber (80 cm) segments in both arms. The spectral filters (F1 and F2) shape the pulse to a narrow-band and short-duration pulse that seeds the propagation in the subsequent arm (Fig. \ref{simulation} (a)). Over the course of its evolution, spectral breathing by a factor of 16 is observed. The pulse evolves quickly to a parabolic shape in the passive fiber (Fig. \ref{simulation} (b)). This parabolic pulse is subsequently amplified in the gain fiber. The parabolic shape is maintained through this gain fiber and into the following passive fiber. This is in contrast to regenerative similariton lasers, where the self-similar evolution is localized to the gain fiber \cite{Finot05, Northa15}. The output in the time domain is a nearly linearly-chirped parabola (Fig. \ref{simulation} (c)) with 110-nm bandwidth (Fig. \ref{simulation} (d)), which corresponds to a $\sim$ 30-fs transform-limited pulse.

Experiments were performed with guidance from the simulations. Yb-doped, PM double-clad fiber with 6-$\mu $m core is employed in the gain segments The 2.5-m long gain segments support the parabolic evolution and absorb most of the pump light. All the passive fibers are standard PM-980. The oscillator's repetition rate is $\sim$17 MHz. The separation of the two filters was adjusted to eliminate continuous-wave (CW) operation while allowing for the highest output pulse energy. With the seed pulses launched into the cavity, the optimal mode-locking conditions were found by adjusting the output coupling in each arm with the waveplates. The seed pulses can be quite weak and their duration is not important. For example, reliable starting was obtained with 80-pJ and 10-ps pulses,  or with \textless10 pJ and 3-ps pulses with 20-nm bandwidth. The bandwidth of the seed pulse is a significant factor because it determines whether a seed can circulate and be amplified in the first roundtrip; if the bandwidth is not wide enough, larger seed energy or higher gain is needed to provide enough spectral broadening. Once pulsation is initiated, as indicated by a broad output spectrum, the seed pulse can be blocked and pulses will continue to circulate in the cavity. While the oscillator is running, physical perturbations can be applied to the fiber without altering the operating state of the laser. Once the optimal conditions are found, oscillation can be extinguished by blocking the cavity or turning off the pump, and restarted to the same state by launching in seed pulses without any additional adjustment. Moreover, an isolated seed pulse is not necessary, as the pulsed operation is stable in the presence of continuous seeding (we tried different continuously-running pulsed lasers as seed sources, operating at 1 MHz and 40 MHz). 

\begin{figure}[htbp]
\centerline{\includegraphics[width=0.7\columnwidth]{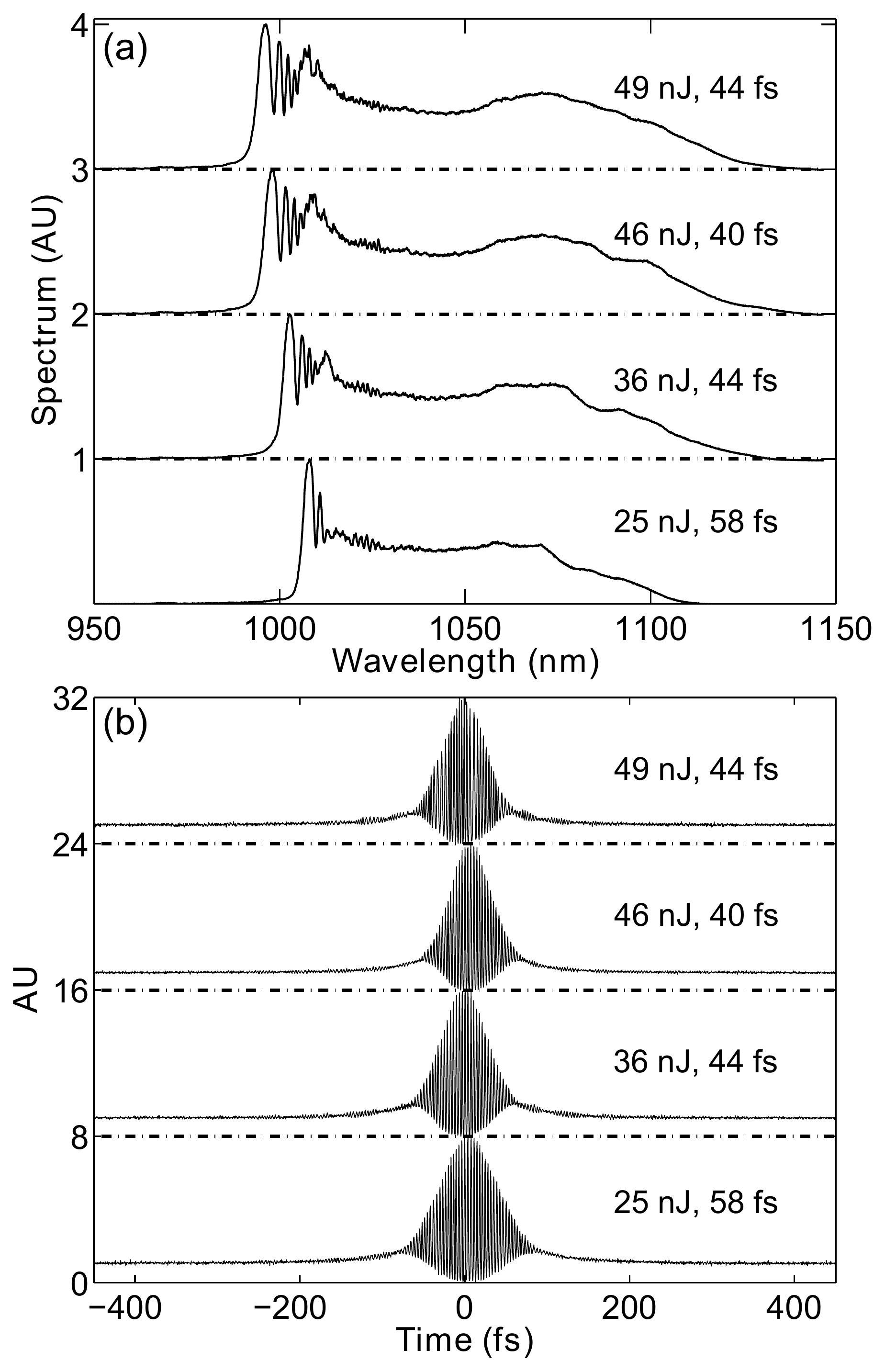}}
\caption{Measurements of pulses from the ring Mamyshev oscillator. (a) Measured output spectra and (b) autocorrelations for the indicated output energies.}
\label{experiment}
\end{figure}

The oscillator generates 6-ps chirped pulses. The output spectra and autocorrelation traces of the dechirped pulses for a range of pulse energies are shown in Fig. \ref{experiment}. Here the output is taken from the output coupler directly before the isolator, and the energy is modified by changing the pump power. As the energy increases, the spectrum broadens due to stronger SPM  (Fig. \ref{experiment} (a)), and the dechirped duration decreases. We find that, using only a grating compressor (300 lines/mm), the dechirped pulses appear relatively clean without pedestals or structure (Fig. \ref{experiment} (b)). For the energy range shown, the dechirped pulse duration is within 1.5 times the transform limit. This deviation can be accounted for by the third-order dispersion in the grating compressor. We estimate that the 50-nJ pulses accumulate a nonlinear phase of 60$\pi$. This shows that the huge nonlinear phase accumulation in the Mamyshev oscillator is well-controlled: it is converted into a nearly-linear chirp. Eventually, higher-order phase that cannot be compensated by the grating pair becomes significant and the minimum dechirped duration grows despite increasing bandwidth (50-nJ trace). The $\sim$50-nJ pulses are limited by the maximum available pump power; we do not observe multi-pulsing, which commonly limits the pulse energy in mode-locked lasers.

\begin{figure}[htbp]
\centerline{\includegraphics[width=1\columnwidth]{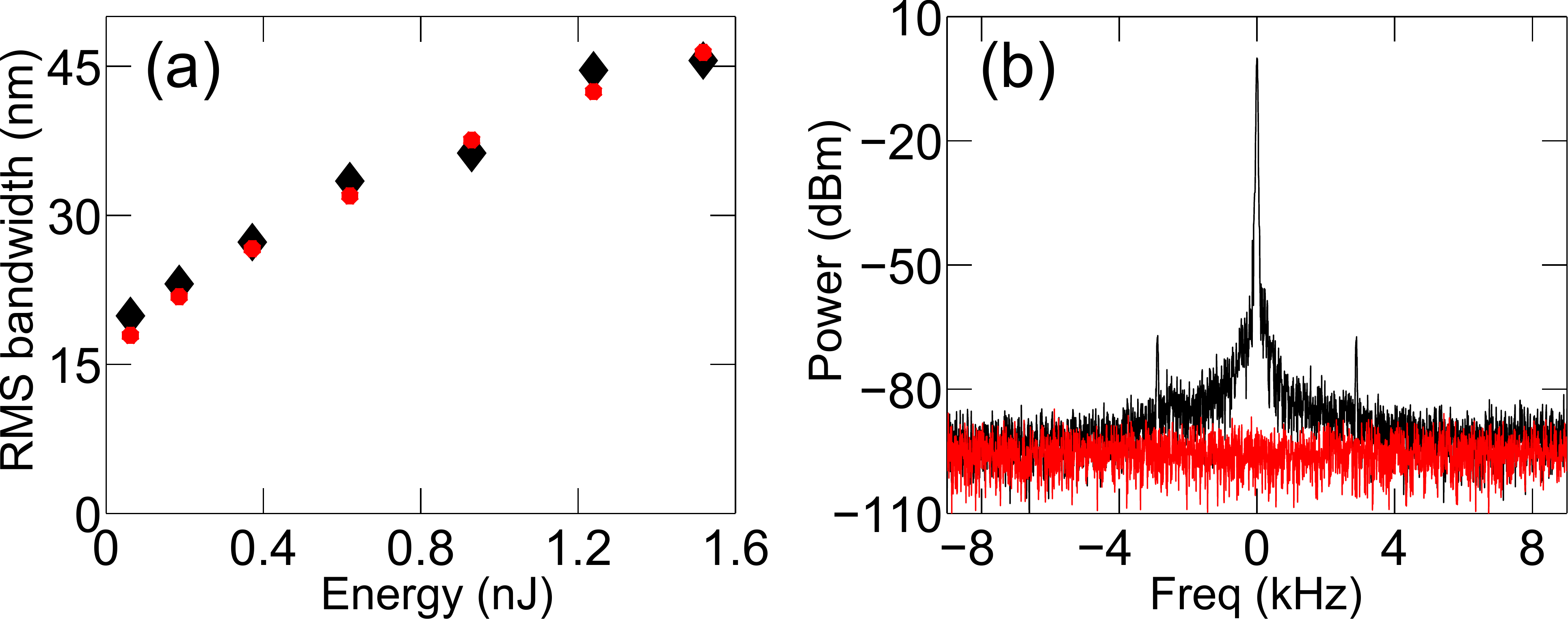}}
\caption{Pulse quality check: (a) measured root-mean-square bandwidth after propagation through 2-m of SMF (black) compared to simulation (red). (b) Radio frequency spectrum with a resolution bandwidth of 30 Hz and a span range of 20 kHz. Noise floor is shown in red.}
\label{sanitycheck}
\end{figure}

The pulse peak power is verified by launching the dechirped pulse into 2-m of SMF with a 6 $\mu$m core diameter (HI1060) and measuring the SPM-induced spectral broadening. The measurements are compared with results of numerical simulations in Fig.\ref{sanitycheck} (a)). In simulation, we launch a Gaussian pulse with the same energy and transform-limited duration, and with the residual third-order dispersion from the grating compressor, into a fiber with the parameters of 2-m of HI1060. The calculation accounts for fiber dispersion up to the third order, SPM, and intrapulse Raman scattering. The simulation reproduces the root-mean-square bandwidth observed for the experimental pulses accurately, which indicates the high quality of the output pulses.

The stability of the output pulse train was investigated using an RF spectrum analyzer. The resolution and dynamic range of the spectra are instrument-limited, but still confirm the stable mode locking and absence of sidebands and harmonic frequencies to at least 80 dB below the fundamental frequency (Fig.\ref{sanitycheck} (b)). One might expect that a regenerative oscillator with significant breathing would exhibit significant pulse-to-pulse fluctuations. We suggest that the self-seeding nature of the Mamyshev oscillator underlies the good stability.

For the conditions described above, the oscillator does not start from noise. Self-pulsation originating from amplified spontaneous emission (ASE) has been predicted \cite{Pitois08} and demonstrated in long cavities ($\sim$km) with highly nonlinear fibers (HNLF) \cite{Sun09,North14}. Starting is favorable in these cavities owing to the narrow filter separation, along with the possibility for sufficient nonlinear phase accumulation by even low-power fluctuations in the long HNLF. We speculate that the recently-described dissipative Faraday instability (DFI) can account for the self-starting in this case, since the small separation between the filters overlaps with the DFI gain spectrum \cite{Perego16,Tarasov16}. For broadband, high-energy pulses, the optimal filter separation is much broader than the DFI gain spectrum. In this regime, self-starting was proposed \cite{RegelskisPatent} and reported \cite{Julijanas15} by use of controlled feedback of amplified spontaneous emission in the linear cavity, through a so-far unexplained mechanism. We have recently confirmed this self-starting in our laboratory (see Supplement 1, section 3). As will be discussed below, although the nature of the Mamyshev oscillator suggests that it is incompatible with self-starting, these initial results provide optimism regarding the near-future realization of a fully self-starting, environmentally-stable fiber oscillator with the high performance demonstrated here. 

\begin{table*}[t]
\centering
\caption{\bf Performance Summary of SMF-based Ytterbium-doped Fiber Oscillator for Different Pulse Evolutions.}
\def\arraystretch{1.6}\tabcolsep=10pt
\begin{tabular}{>{\centering \arraybackslash} p{3 cm}  >{\centering \arraybackslash}  p{3cm} >{\centering \arraybackslash}  p{3 cm}>{\centering \arraybackslash}  p{5cm}}
\hline
 Pulse evolution & Nonlinear phase  &  Typical performance  & Best performance \\
\hline
Soliton & $\sim$ 0 &0.1 nJ, 300 fs &0.5 nJ, 100 fs \\
\hline
Stretched pulses  & 0 to $\pi$&1 nJ, 100 fs	 &  up to 3 nJ, down to 50 fs\\
\hline
Passive similariton &2$\pi$ - 10$\pi$ & 6 nJ, 150 fs  &15 nJ, 100 fs \\
\hline
Dissipative soliton  & 2$\pi$ - 10$\pi$  &6 nJ, 150 fs    & up to 20 nJ, down to 70 fs \\
\hline
Amplifier similariton  &4$\pi$ - 10$\pi$ &3 nJ, 70 fs	 & up to  8 nJ, 40 fs \\
\hline
Ti:sapphire &0 to $\pi$&30 nJ, 50-100 fs  &  200 nJ, 30 fs \\
\hline
Mamyshev oscillator  & \textgreater60$\pi$   &   & experiment: 50 nJ, 40 fs  \hspace{1cm} (in simulation: \textgreater190 nJ, \textless20 fs) \\
\hline
\end{tabular}
  \label{differentevolution}
\end{table*}

\section{Discussion}
The performance of mode-locked lasers is fundamentally limited by nonlinear effects. Solitons and dispersion-managed solitons are stable for at most a round-trip peak nonlinear phase shift of $\pi$. Experiments showed the passive similariton, dissipative soliton and amplifier similariton can support \textless100 fs and \textgreater10 nJ pulses, which corresponds to $\sim$ 10$\pi$ nonlinear phase shift \cite{Renninger2015,Chong2015}. Simulation of these evolutions, assuming ideal saturable absorbers, indicate that higher performance can be achieved with higher pump power. These high-energy pulses, \textit{i.e.} larger nonlinear phase shift, require a very high modulation-depth saturable absorber to suppress the CW background \cite{Renninger2015}. While NPE can be close to an ideal absorber, there is still a significant gap between simulations and experiments, and currently $\sim$ 10$\pi$ represents an approximate limit for experiments. Table \ref{differentevolution} summarizes the performance of representative Yb-doped mode-locked fiber lasers.

The Mamyshev oscillator overcomes these limitations. If the nonlinearity is correctly managed, high energy, wave-breaking-free pulses with good phase profile can be generated - apparently even well beyond the gain bandwidth limit. This performance follows from the ``perfect'' saturable absorber realized by the combination of two filters and the fiber. As pointed out by Pitois \textit{et al.}, the cascaded frequency-broadening and offset filtering creates an effective transmission function that is step-like, with zero transmittance at low power and an abrupt transition to a constant value at high power \cite{Pitois08}. This perfect saturable absorber means that the mode-locking pumping rate is below the continuous wave lasing threshold, so the Mamyshev oscillator only supports mode-locked operation. This eliminates the nonlinearity limit from the saturable absorber \cite{Renninger2015}, and so allows for much higher energy. Already, the compressed pulse we obtain has a peak power roughly an order-of-magnitude above previous results. The performance is comparable that of the best commercial Ti:sapphire lasers, the current workhorses of ultrafast laser applications (Table \ref{differentevolution}).

Of course, the ``perfect" saturable absorber property creates a challenge for starting a Mamyshev oscillator from noise. Despite this, two methods of starting similar cavities have recently been demonstrated: pump modulation \cite{Samartsev17} and controlled feedback of amplified spontaneous emission \cite{Julijanas15}. Our own initial investigations confirm that the latter is, remarkably, a reliable and robust means of circumventing the starting problem (see Supplement 1, section 3). While the pump modulation scheme demonstrated by Samartsev et al. \cite{Samartsev17} is important, using a coupled cavity to start the Mamyshev oscillator has the practical benefit of being completely passive. It is surprising that any starting technique can be successful, because of the inevitable trade-off between starting and performance and the bias here towards performance due to the ``perfect'' saturable absorber property. That researchers have demonstrated such methods in similar cavities suggests that related techniques may be used for the ring cavity presented here. Future work will be focused on developing such techniques.

Prospects for further optimization and scaling of the Mamyshev oscillator are exciting. While simulations of ultrafast fiber sources have systematically suggested much higher performance than has been observed, we are more optimistic about our extrapolations for the Mamyshev oscillator. The discrepancy between experimental performance and anticipated results in conventional fiber lasers likely follows from uncertain starting conditions and overly-optimistic saturable absorber parameters. Starting the oscillator from a pulse systematically produces higher performance than starting from noise \cite{Bucklew2017}. In contrast, our observations in the self-starting linear Mamyshev oscillator show that pulse-seeded and self-starting performance are similar. Meanwhile, standard numerical techniques for modeling ultrashort-pulse fiber lasers neglect gain relaxation dynamics and use a restricted temporal window. These fail to account for nanosecond laser spiking or other effects that may play an important role in starting, and ultimately the steady-state performance. In the Mamyshev oscillator, CW lasing is completely suppressed: the CW lasing threshold is much \textit{higher} than the pulse threshold. Consequently, if achieved, starting should be more deterministic, and the maximal nonlinear phase shift can be much higher than in a cavity where CW lasing needs to be constantly suppressed. 

Scaling from the experimental results presented here, with large mode area (25 $\mu$m diameter) PM fiber we expect that Mamyshev oscillators will reach the microjoule level. Simulations also indicate that by changing the fiber length, the oscillator repetition rate can be tuned from hundreds of MHz to $\sim$ 1 MHz without sacrificing the performance (at lower repetition rates, additional dispersion compensation inside of the cavity is required). This will allow it to be a useful tool for both scientific and industrial applications. In addition, the linearly-chirped parabolic output pulses are attractive for applications such as highly-coherent continuum sources and optical signal processing, etc \cite{Finot09}. 

Given that it has taken nearly a decade for the Mamyshev pulse generator to gain serious consideration as an alternative to conventional mode-locked fiber lasers, it is worthwhile to reconsider broader impacts of this system. Numerous works \cite{Provost07, Pitois08} have explored optimization of the Mamyshev oscillator for signal regeneration. The results presented here explore the system in connection and contrast to mode-locked fiber lasers. Much work remains to optimize oscillator designs for this purpose. However, a crucial difference between the Mamyshev oscillator and a conventional mode-locked laser is that the pulsed state is bistable with the continuous-wave state (or ASE). Hence the Mamyshev oscillator should be compared and contrasted not only with mode-locked lasers, but also with systems that support cavity solitons, such as coherently-driven cavities, which have lately been explored extensively for producing mode-locked frequency combs \cite{Leo2010,Herr20103,Chembo2013,Coen2013}. Two features of the pulse evolution we demonstrate here are worth noting. First, the Mamyshev oscillator can produce a stable pulse train even with extremely high roundtrip nonlinear phase shift and spectro-temporal breathing. This suggests that in a suitably-designed device, an octave-spanning frequency comb could be generated directly, possibly exceeding the bandwidth and certainly the power of microresonator combs. Second, as in fiber lasers, the roundtrip gain and loss are much higher than in coherently-driven cavities. This may allow more straightforward control over circulating pulses, at the point of minimum pulse energy by, for example, an intracavity electro-optic modulator or coupling to an external source of optical bits. Furthermore, following from the possibility for extreme nonlinear phase and pulse evolution, much more elaborate information processing schemes could be devised. Ultimately, these suggestions are only speculative examples of the more important message. Many questions remain about the Mamyshev oscillator, and its unique features suggest a wide variety of uses and phenomena which, to date, have been under-explored compared to conventional mode-locked laser cavities and coherently-driven high-Q optical resonators.

\section{Conclusion}

In conclusion, these results show that the Mamyshev oscillator allows a surprising and significant leap in the ongoing central challenge of high-power ultrafast fiber lasers: the management of nonlinearity. This is due to the formation and amplification of parabolic pulses and the ``perfect'' artificial saturable absorber formed by the Mamyshev regeneration mechanism. By making the continuous wave lasing threshold above the threshold for mode-locking, the Mamyshev oscillator supports stable mode-locking with huge nonlinear phase shifts. To harness this nonlinearity for the generation of clean, high-energy ultrashort pulses, we designed the oscillator to support parabolic pulse formation. All these translate directly to unprecedented performance - our initial design already yields an order-of-magnitude higher peak power than any previous fiber oscillator with the same core size. As prior work has shown, the Mamyshev oscillator supports an environmentally-stable, fiber-format design which can be self-starting, thus solving a major practical impediment to the widespread use of ultrashort-pulse fiber sources in applications. Taken together, these features should make Mamyshev oscillator extremely attractive for applications in ultrafast science and technology. 

\section*{Funding Information}
Portions of this work were funded by the Office of Naval Research (N00014-13-1-0649) and the National Institutes of Health (EB002019). Z. Ziegler acknowledges support from an ELI undergraduate research award at Cornell University.

\section*{Acknowledgments}
The authors gratefully acknowledge helpful discussions with J. Zeludevicius.  

\section{Supplementary Info}
See Supplement 1 for supporting content.


\begin{thebibliography}{0}%
\makeatletter
\providecommand \@ifxundefined [1]{%
 \@ifx{#1\undefined}
}%
\providecommand \@ifnum [1]{%
 \ifnum #1\expandafter \@firstoftwo
 \else \expandafter \@secondoftwo
 \fi
}%
\providecommand \@ifx [1]{%
 \ifx #1\expandafter \@firstoftwo
 \else \expandafter \@secondoftwo
 \fi
}%
\providecommand \natexlab [1]{#1}%
\providecommand \enquote  [1]{``#1''}%
\providecommand \bibnamefont  [1]{#1}%
\providecommand \bibfnamefont [1]{#1}%
\providecommand \citenamefont [1]{#1}%
\providecommand \href@noop [0]{\@secondoftwo}%
\providecommand \href [0]{\begingroup \@sanitize@url \@href}%
\providecommand \@href[1]{\@@startlink{#1}\@@href}%
\providecommand \@@href[1]{\endgroup#1\@@endlink}%
\providecommand \@sanitize@url [0]{\catcode `\\12\catcode `\$12\catcode
  `\&12\catcode `\#12\catcode `\^12\catcode `\_12\catcode `\%12\relax}%
\providecommand \@@startlink[1]{}%
\providecommand \@@endlink[0]{}%
\providecommand \url  [0]{\begingroup\@sanitize@url \@url }%
\providecommand \@url [1]{\endgroup\@href {#1}{\urlprefix }}%
\providecommand \urlprefix  [0]{URL }%
\providecommand \Eprint [0]{\href }%
\providecommand \doibase [0]{http://dx.doi.org/}%
\providecommand \selectlanguage [0]{\@gobble}%
\providecommand \bibinfo  [0]{\@secondoftwo}%
\providecommand \bibfield  [0]{\@secondoftwo}%
\providecommand \translation [1]{[#1]}%
\providecommand \BibitemOpen [0]{}%
\providecommand \bibitemStop [0]{}%
\providecommand \bibitemNoStop [0]{.\EOS\space}%
\providecommand \EOS [0]{\spacefactor3000\relax}%
\providecommand \BibitemShut  [1]{\csname bibitem#1\endcsname}%
\let\auto@bib@innerbib\@empty
\end{thebibliography}%


\begin{thebibliography}{99}

\bibitem{Ilday04}
F. O. Ilday, J. R. Buckley, W. G. Clark, and F. W. Wise, ``Self-Similar Evolution of Parabolic Pulses in a Laser,'' Phys. Rev. Lett. \textbf{92}, 213902 (2004).

\bibitem{Chong06}  A. Chong, J. Buckley, W. Renninger, and F. W. Wise, ``All-normal-dispersion femtosecond fiber laser,'' Opt. Express \textbf{14}, 10095–10100 (2006).

\bibitem{Chong20nj07}A. Chong, W. H. Renninger, and F. W. Wise, ``All-normal-dispersion femtosecond fiber laser with pulse energy above 20 nJ,'' Opt. Lett. \textbf{32}, 2408-2410 (2007).

\bibitem{Nie11}
B. Nie, D. Pestov, F. W. Wise, and M. Dantus, ``Generation of 42-fs and 10-nJ pulses from a fiber laser with self-similar evolution in the gain segment,'' Opt. Express \textbf{19}, 12074-12080 (2011).


\bibitem{Chong2015} A. Chong, L. G. Wright, and F. W. Wise, ``Ultrafast fiber lasers based on self-similar pulse evolution: a review of current progress,'' Rep. Prog. Phys. \textbf{78}, 113901 (2015).


\bibitem{Moll1984} L. F. Mollenauer and R. H. Stolen, ``The soliton laser,'' Opt. Lett. \textbf{9}, 13-15 (1984).

\bibitem{Tamura1993} K. Tamura, E. P. Ippen, H. A. Haus, and L. E. Nelson, ``77-fs pulse generation from a stretched-pulse mode-locked all-fiber ring laser,'' Opt. Letters \textbf{18}, 1080-1082 (1993).

\bibitem{Keller96}
U. Keller, K. J. Weingarten, F. X. Kärtner, D. Kopf, B. Braun, I. D. Jung, R. Fluck, C. Hönninger, N. Matuschek, and J. Aus der Au, "Semiconductor Saturable Absorber Mirrors (SESAM's) for femtosecond to nanosecond pulse generation in solid-state lasers," IEEE J. Sel. Top. Quantum Electron. \textbf{2}, 435-453 (1996).

\bibitem{Chong08} A. Chong, W. H. Renninger, and F. W. Wise, ``Environmentally stable all-normal-dispersion femtosecond fiber laser,'' Opt. Lett. \textbf{33}, 1071-1073 (2008).

\bibitem{Aguergaray12}
C. Aguergaray, N. G. R. Broderick, M. Erkintalo, J. S. Y. Chen, and V. Kruglov, ``Mode-locked femtosecond all-normal all-PM Yb-doped fiber laser using a nonlinear amplifying loop mirror,'' Opt. Express \textbf{20}, 10545-10551 (2012).

\bibitem{Aguergaray13}
C. Aguergaray, R. Hawker, A. F. J. Runge, M. Erkintalo, and N. G. R. Broderick, ``120 fs, 4.2 nJ pulses from an all-normal-dispersion, polarization-maintaining, fiber laser,'' Appl. Phys. Lett. \textbf{103}, 121111 (2013).

\bibitem{Szczepanek15}
J. Szczepanek, T. M. Kardaś, M. Michalska, C. Radzewicz, and Y. Stepanenko, ``Simple all-PM-fiber laser mode-locked with a nonlinear loop mirror,'' Opt. Lett. \textbf{40}, 3500-3503 (2015).

\bibitem{Pitois08}
S. Pitois, C. Finot, L. Provost, and D. J. Richardson, `` Generation of localized pulses from incoherent wave in optical fiber lines made of concatenated Mamyshev regenerators,'' J. Opt. Soc. Am. B \textbf{25}, 1537-1547 (2008)

\bibitem{Sun09}
K. Sun, M. Rochette, and L. R. Chen, ``Output characterization of a self-pulsating and aperiodic optical fiber source based on cascaded regeneration,'' Opt. Express \textbf{17}, 10419-10432 (2009).

\bibitem{North14}
T. North and M. Rochette, ``Regenerative self-pulsating sources of large bandwidths,'' Opt. Lett. \textbf{39}, 174-177 (2014).

\bibitem{Mamyshev98}
P. V. Mamyshev, ``All-optical data regeneration based on self-phase modulation effect,'' in Proceedings of 24th European Conference on Optical Communication, Madrid, Spain (IEEE, 1998), p. 475-476.

\bibitem{Provost07}
L. Provost, C. Finot, P. Petropoulos, K. Mukasa, and D. J. Richardson, ``Design scaling rules for 2R-optical self-phase modulation-based regenerators,'' Opt. Express \textbf{15}, 5100–5113 (2007).

\bibitem{Regelskis15} 
K. Regelskis, J. Zeludevicius, K. Viskontas, and G. Raciukaitis, ``Ytterbium-doped fiber ultrashort pulse generator based on self-phase modulation and alternating spectral filtering,'' Opt. Lett. \textbf{40}, 5255-5258 (2015).

\bibitem{RegelskisPatent}
K. Regelskis, G. Raciukaitis, ``Method and generator for generating ultra-short light pulses,'' WO2016020188 A1.

\bibitem{Samartsev17}
I. Samartsev, A. Bordenyuk, V. Gapontsev, ``Environmentally stable seed source for high power ultrafast laser,'' Proc. SPIE 10085, Components and Packaging for Laser Systems III, 100850S (February 22, 2017); doi:10.1117/12.2250641.

\bibitem{Renninger2010}
W. H. Renniger, A. Chong, and F. W. Wise, ``Self-similar pusle evolution in an all-normal-dispersion laser,", Phys. Rev. A \textbf{82}, 021805(R) (2010).

\bibitem{Finot05}
C. Finot, S. Pitois, and G. Millot, ``Regenerative 40 Gbit/s wavelength converter based on similariton generation,'' Opt. Lett. \textbf{30}, 1776–1778 (2005).

\bibitem{Northa15}
T. Northa and C. S. Bres, ``Regenerative similariton laser,'' APL Photonics \textbf{1}, 021302 (2016).

\bibitem{Perego16}
A. M. Perogo, N. Tarasov, D. V. Churkin, S. K. Turisyn, and K. Staliunas, ``Pattern Generation by Dissipative Parametric Instability,'' Phys. Rev. Lett \textbf{116}, 028701 (2016).

\bibitem{Tarasov16}
N. Tarasov, A. M. Perego, D. V. Churkin, K. Staliunas, and S. K. Turitsyn, ``Mode-locking via dissipative Faraday instability,'' Nat. Commun. \textbf{7}, 12441 (2016).

\bibitem{Julijanas15}
Personal communication with J. Zeludevicius, Center for Physical Sciences \& Technology (CPST) Savanoriu Ave. 231, LT-02300 Vilnius, Lithuania (2015).



\bibitem{Renninger2015}
W. H. Renninger and F. W. Wise, ``Fundamental Limits to Mode-Locked Lasers: Toward Terawatt Peak Powers,'' IEEE J. Sel. Top. Quantum Electron. \textbf{21}, 1100208 (2015).

\bibitem{Bucklew2017}
V. G. Bucklew, W. H. Renninger, P. S. Edwards, Z. Liu, "Iteratively seeded mode-locking," arxiv: 1612.04269 (2017)


\bibitem{Finot09}
C. Finot, J. Dudley, B. Kibler, D. Richardson, and G. Millot, ``Optical Parabolic Pulse Generation and Applications,'' IEEE J. Quantum Electron. \textbf{45}, 1482–1489 (2009).

\bibitem{Leo2010}
F. Leo, S. Coen, P. Kockaert, S.-P. Gorza, P. Emplit, and M. Haelterman, ``Temporal cavity solitons in one-dimensional Kerr media as bits in an all-optical buffer,'' Nature Photon. \textbf{4}, 471-476 (2010).

\bibitem{Herr20103}
T. Herr, V. Brasch, J. D. Jost, C. Y. Wang, N. M. Kondratiev, M. L. Gorodetsky, and T. J. Kippenberg, ``Temporal solitons in optical microresonators,'' Nature Photon. \textbf{8}, 145-152 (2014).

\bibitem{Chembo2013}
Y. K. Chembo and C. R. Menyuk, ``Spatiotemporal Lugiato-Lefever formalism for Kerr-comb generation in whispering-gallery-mode resonators,'' Phys. Rev. A \textbf{87}, 053852 (2013).

\bibitem{Coen2013}
S. Coen and M. Erkintalo, ``Universal scaling laws of Kerr frequency combs,'' Opt. Lett. \textbf{38}, 1790-1792 (2013)


\end{thebibliography}
\end{document}


\title{Megawatt peak power from a Mamyshev oscillator: supplementary material}

\author{Zhanwei Liu}\email{Corresponding author: zl358@cornell.edu}
\author{Zachary M. Ziegler}
\author{Logan G. Wright}
\author{Frank. W. Wise}
\affiliation{School of Applied and Engineering Physics, Cornell University, Ithaca, New York 14853, USA}

\begin{abstract}
This document provides supplementary information to \textit{Megawatt peak power from a Mamyshev oscillator}. We show numerical results on maximum performance of the experimental Mamsyshev oscillator design and discuss performance limits. We provide numerical results that show the importance of a Gaussian filter profile for obtaining high-quality, high-energy pulses. Finally, experiments on starting a linear cavity are described.
\end{abstract}

\ocis{(060.2310) Fiber optics; (190.0190) Nonlinear optics; (060.4370)  Nonlinear optics, fibers.}

\maketitle 

\section{Simulation result for the highest energy pulses}

\begin{figure}[h]
\centerline{\includegraphics[width=1\columnwidth]{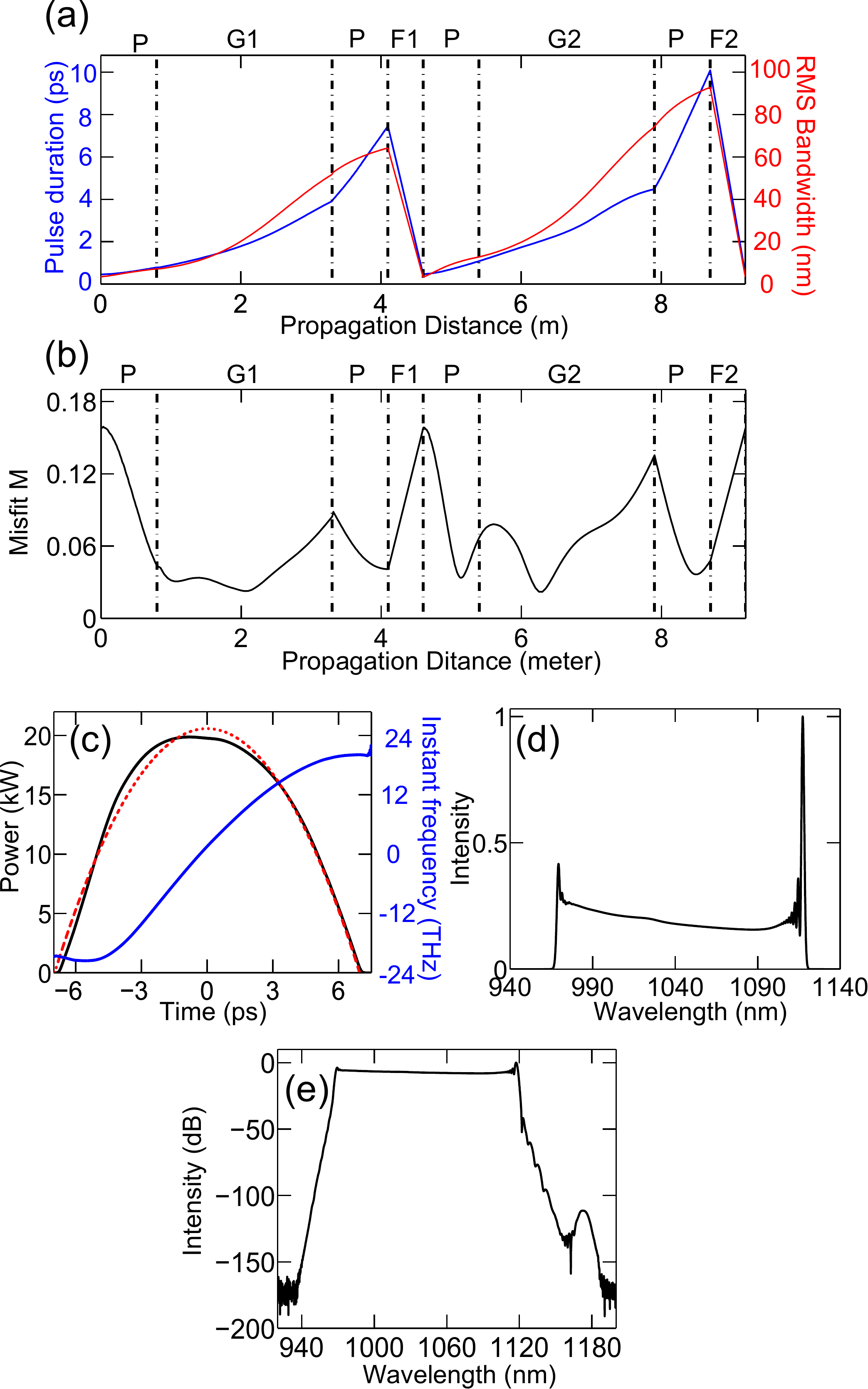}}
\caption{Numerical simulation results for 190 nJ pulses. (a) Evolution of pulse duration (blue) and RMS bandwidth (red), P: passive fiber; G: gain fiber; F: filter; (b) evolution of misfit parameter M, $M^2 = \int (I-I_{fit})^2 dt / \int I^2 dt$, which indicates the difference between the pulse shape ($I$) and its parabolic fitting profile $I_{fit}$; (c) temporal profile (black) with fitted parabolic curve (red) and instantaneous frequency across the chirped pulse (blue) and (d) (linear scale) and (e) (decibel scale) of the simulated spectrum. }
\label{190nJSim}
\end{figure}

Fig. \ref{190nJSim} shows the simulation result for the highest pulse energy output (190 nJ) for the same cavity design as the experiment. The spectral breathing ratio is 26 (Fig. \ref{190nJSim} (a)). Compared with the 50-nJ example, the pulse shape deviates more from the parabola during the whole roundtrip (Fig. \ref{190nJSim} (b)). The chirp of the pulse is consequently not linear, and its deviation increases for higher pulse energy (Fig. \ref{190nJSim} (c)). That said, the pulse is still coherent, and the 150-nm bandwidth (Fig. \ref{190nJSim} (d)) corresponds to \textless20 fs transform-limited pulses, which could be obtained with a pulse compressor that controls higher-order dispersion.  On a logarithmic scale, the spectrum shows the formation of a parasitic Raman peak (Fig. \ref{190nJSim} (e)). Understanding these two factors - breakdown of the parabolic shape, and formation of Raman - and how to manage them will determine the ultimate limits of the Mamyshev oscillator. 

\section{Effect of the filter shape}

\begin{figure}[H]
\centerline{\includegraphics[width=1\columnwidth]{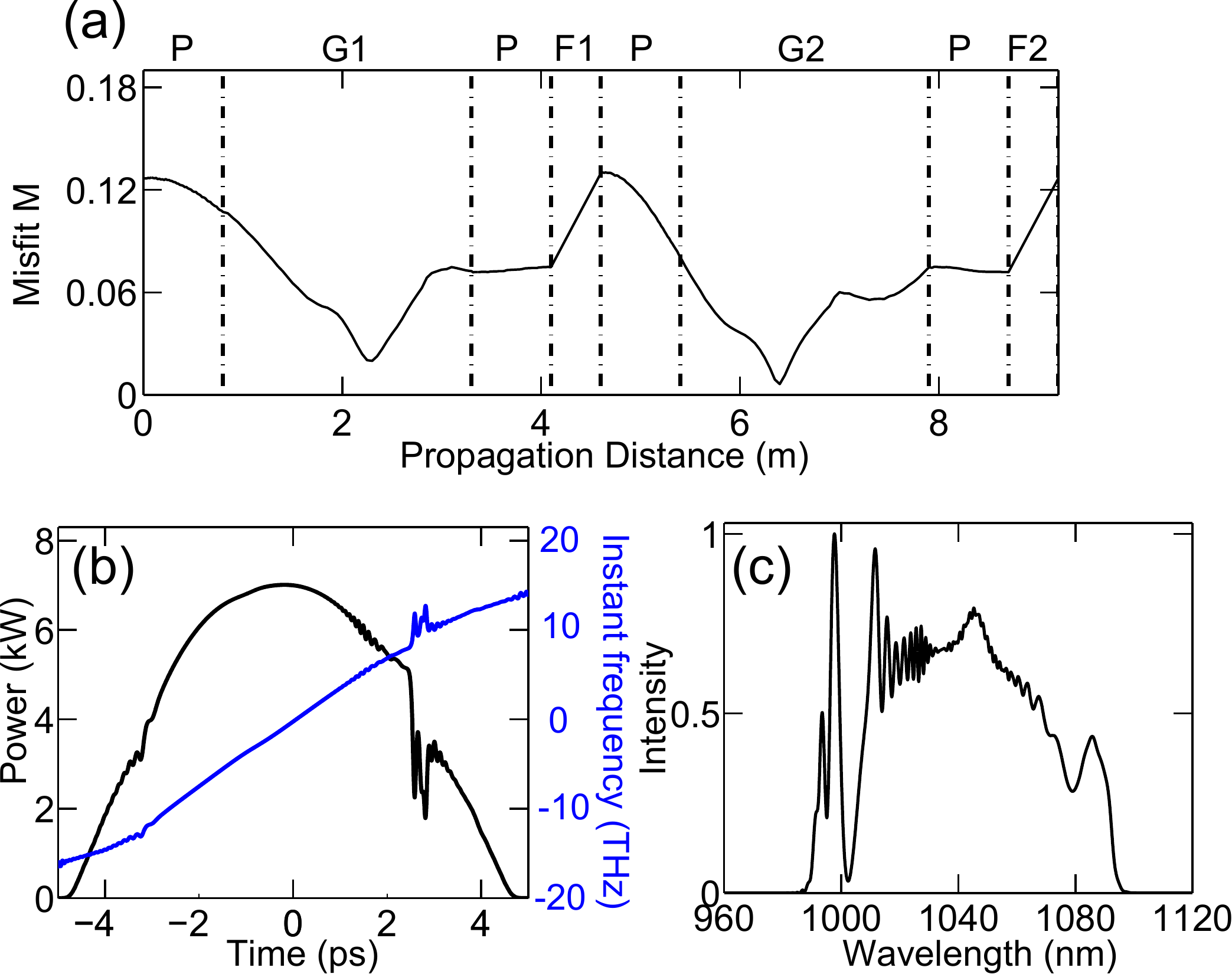}}
\caption{Simulated ring Mamyshev oscillator using super-Gaussian filters.  The pulse energy is 42 nJ. (a) Evolution of misfit parameter M, (b) temporal profile (black) and instantaneous frequency across the chirped pulse (blue) and (c) the simulated spectrum.}
\label{42nJsuperGaussian}
\end{figure}

Using a Gaussian filter is evidently important for the peak performance and pulse quality. For the ring cavity, when we replace the Gaussian filters (4-nm bandwidth) with super-Gaussian shape filters (4-nm bandwidth) in the simulation, the highest achieved pulse energy is $\sim$ 40 nJ, which is much lower than the results using Gaussian filters (Fig. \ref{42nJsuperGaussian}).

\begin{figure}[H]
\centerline{\includegraphics[width=1\columnwidth]{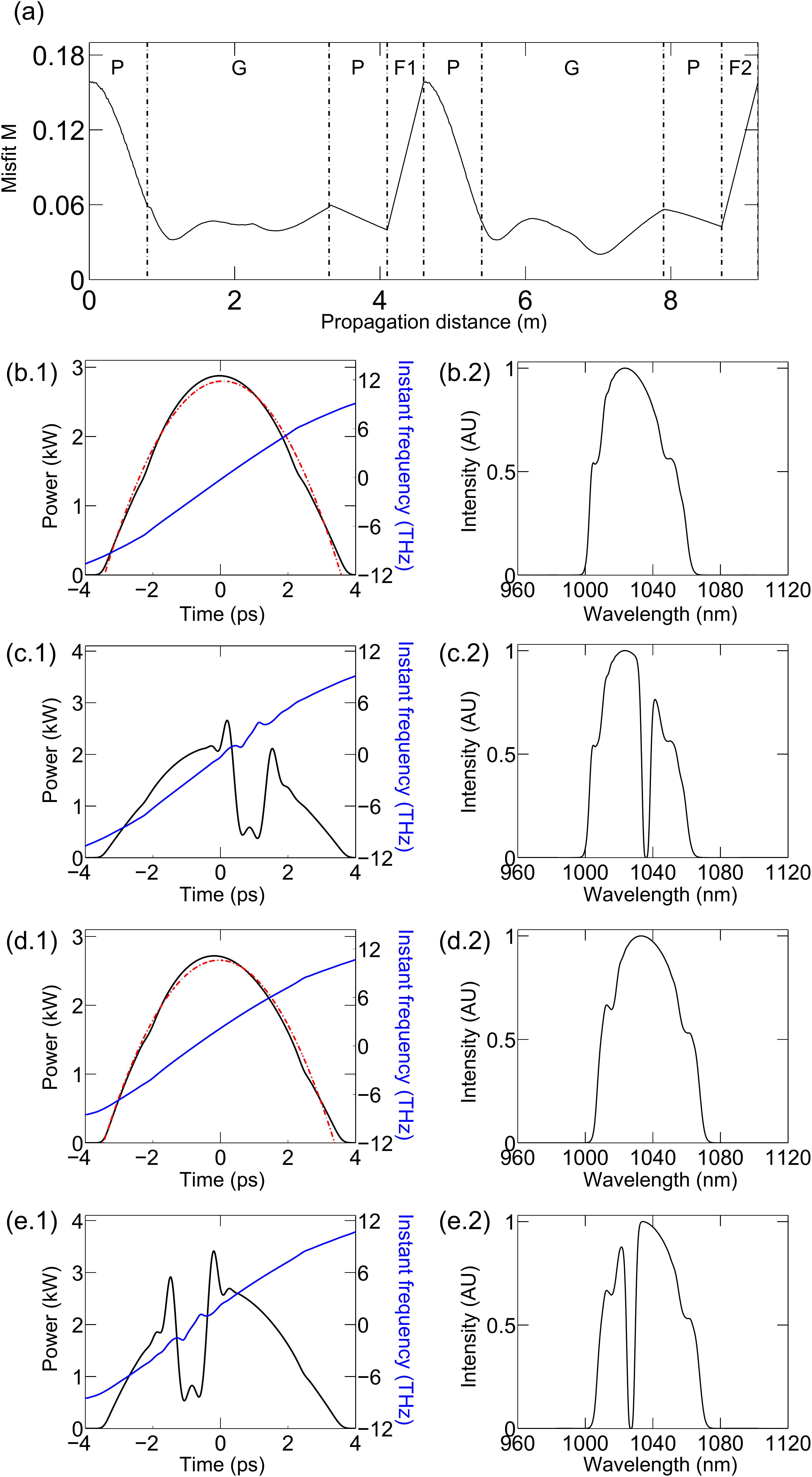}}
\caption{Simulated linear Mamyshev oscillator using Gaussian filters. (a.1, a.2) pulse shape and spectrum before output 1, (b.1, b.2) output pulse shape and spectrum from filter 1, (c.1, c.2) pulse shape and spectrum before output 2 (d.1, d.2) output pulse shape and spectrum from filter 2.}
\label{gaussianfilterLinearcavity}
\end{figure}

\begin{figure}[H]
\centerline{\includegraphics[width=1\columnwidth]{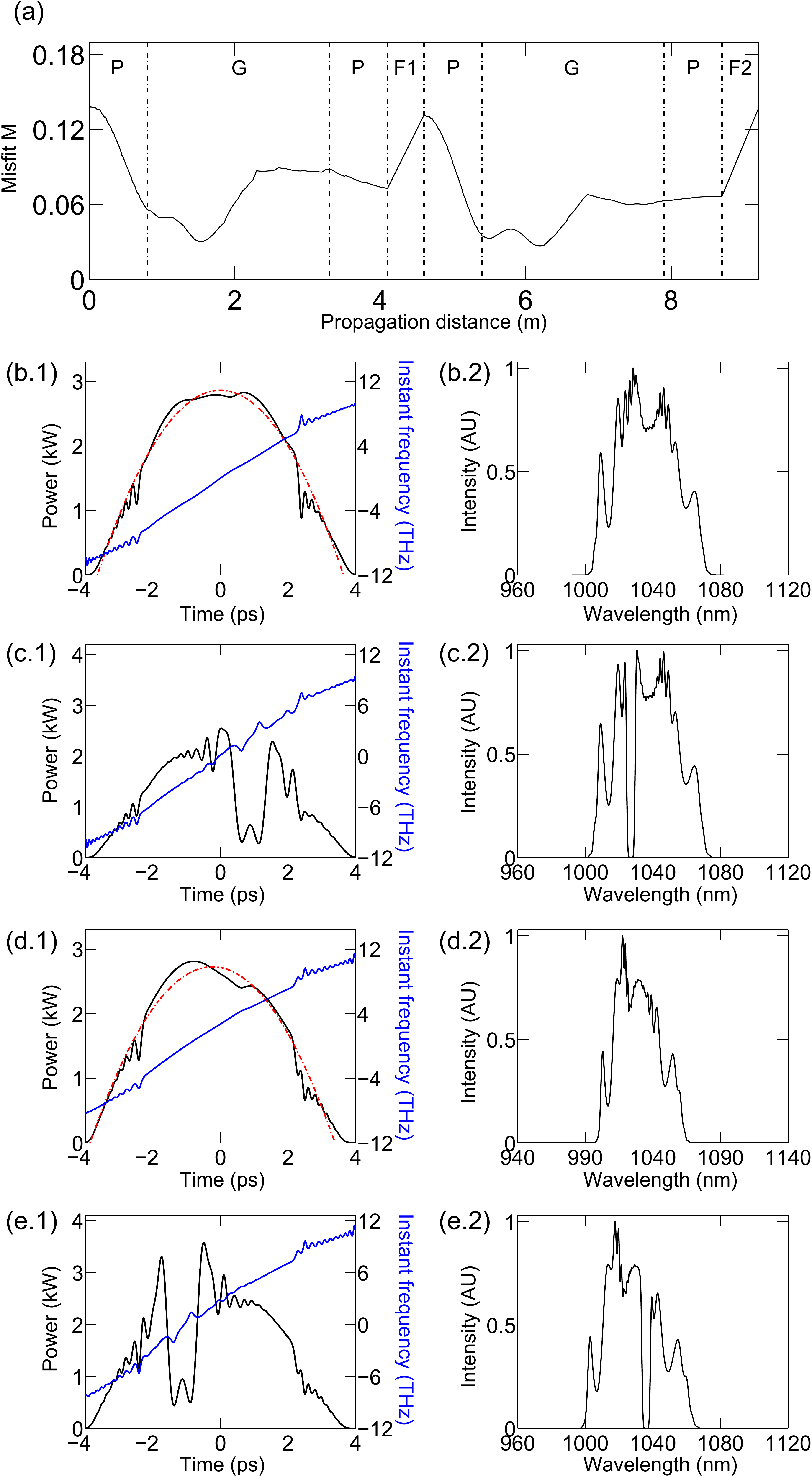}}
\caption{Simulated linear Mamyshev oscillator using super-Gaussian filters. (a.1, a.2) pulse shape and spectrum before output 1, (b.1, b.2) output pulse shape and spectrum from filter 1, (c.1, c.2) pulse shape and spectrum before output 2 (d.1, d.2) output pulse shape and spectrum from filter 2.}
\label{supergaussianfilterLinearcavity}
\end{figure}

The simulations indicate that the use of filters that do not have Gaussian transmittance bands reduces the stable pulse energy and yields structure on the spectrum. Furthermore, the Gaussian filter's improvement of the output chirped pulses' smooth parabolic shape, and the smoothness of the spectra, is an important feature that should improve the pulses' use as a seed for low-distortion fiber amplifiers. Fig. \ref{gaussianfilterLinearcavity} and Fig. \ref{supergaussianfilterLinearcavity} show the difference between the linear cavity output using Gaussian filter and super-Gaussian filter with similar bandwidth (4nm FWHM).

\section{Experiments on starting}
\begin{figure}[h]
\centerline{\includegraphics[width=1\columnwidth]{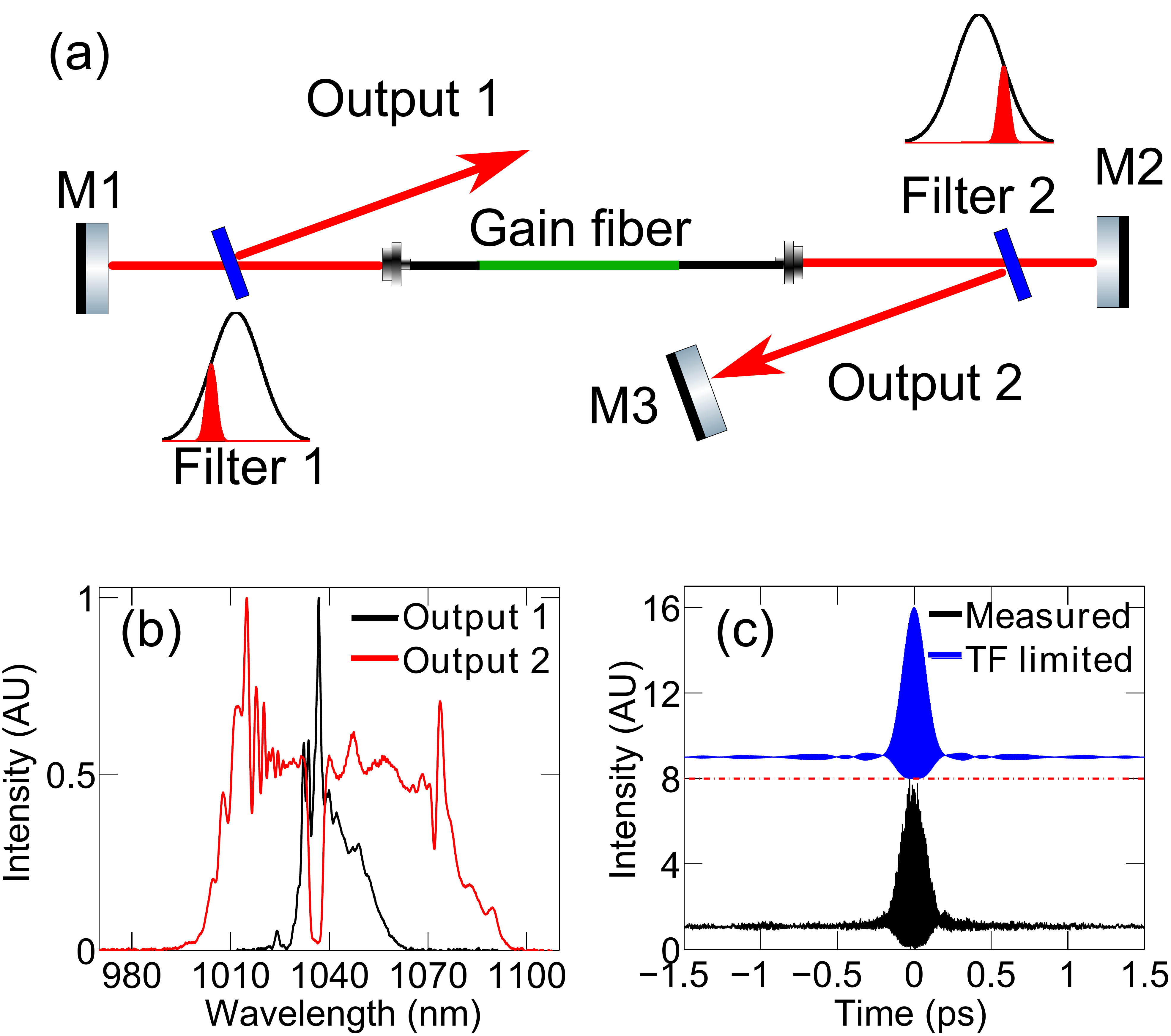}}
\caption{Linear Mamyshev oscillator used to study self-starting. (a) Schematic of the oscillator, (b) measured spectra of the outputs, and (c) transform limited and measured autocorrelation for output 1.}
\label{linear}
\end{figure}

To investigate the starting dynamics, we built a similar linear oscillator (Fig. \ref{linear} (a)). Two mirrors (M1 and M2), two bandpass filters (4-nm near super-Gaussian interference filters biased to longer and shorter wavelengths with respect to the gain spectrum) and the fiber segments form the Mamyshev cavity. The rejected beams from these filters are taken as the outputs. When seeded by short pulses as described in the main text, the linear oscillator behaves like the ring, but the maximum pulse energy is lower due to several practical problems. For example, simulations show that output performance is limited compared to the ring due to the super-Gaussian filter shape. Short-pulse generation can also be started by placing a mirror (M3) at one of the outputs to direct the rejected light back into the oscillator. This is consistent with previously reported work \cite{Julijanas15,RegelskisPatent}. Adding this mirror creates two coupled cavities: the Mamyshev cavity with two spectral filters, described above, and a cavity that can oscillate on a CW basis within the pass-band of filter 1. The Mamyshev cavity only oscillates for high peak power and does not support CW operation, while the opposite is true for the coupled CW cavity formed by M1 and M3. With the feedback cavity coupled, we observe that a pulse train in the Mamyshev typically starts after $\sim$ 10 seconds of operation. The spectra recorded from the two outputs are shown in Fig. \ref{linear} (b), and are very similar to the spectra produced with seed pulses. The spectrum from output 2 (13 nJ) is wider than the spectrum from output 1 (6.5 nJ). The large bandwidth difference between the two outputs is a result of the filters not being symmetrically distant from the peak of the gain spectrum. The measured autocorrelation for the dechirped output 1 (using a grating pair compressor) and the corresponding transform-limited pulses (125-fs duration) are shown in Fig. \ref{linear} (c). The close agreement implies that the chirp of the pulses exiting the cavity must be nearly linear. In contrast, the pulse from output 2 cannot be dechirped to the transform limit due to  uncontrolled nonlinear phase accumulation. This is likely a consequence of the non-Gaussian filter profile.

By observing the cavity output prior to starting, we infer that self-starting in the coupled linear cavity Mamyshev oscillator is the result of spectral broadening of extreme lasing spikes that result from relaxation oscillations within the coupled cavity. We observe bunches of nanosecond-duration pulses from output 1, whose duration and $\sim$ hundred-$\mu$s temporal separation is consistent with relaxation oscillation in the CW cavity. These relaxation oscillations are only observed when the loss of the CW cavity is sufficiently large; when M3 is optimally-aligned, self-starting is not observed. When self-starting is observed, it is repeatable: if the cavity is blocked and unblocked, pulsation returns within $\sim$ 10 seconds. We attribute the self-starting to spectral broadening of the nanosecond-duration spikes. For rare extreme-amplitude spikes, relatively broad initial bandwidth and nonlinear spectral broadening are sufficient for them to couple into the Mamyshev cavity and initiate pulsation. Once the Mamyshev cavity has started, it is indefinitely stable in the continued presence of the coupled CW cavity, which is attractive for practical applications.

